\input harvmac.tex

\def\phi{\varphi}

\Title{quant-ph/0310149}
{\vbox{\centerline{Applications of Partial Supersymmetry}}}

\centerline{Donald
Spector\footnote{$^\dagger$}{spector@hws.edu} }
\medskip\centerline{Department of Physics, Eaton Hall}
\centerline{Hobart and William Smith Colleges}
\centerline{Geneva, NY \ 14456 USA}

\vskip .3in
I examine quantum mechanical Hamiltonians with partial supersymmetry,
and explore two main applications.  First, I analyze a theory with a logarithmic
spectrum, and show how to use partial supersymmetry to reveal the underlying
structure of this theory.  This method reveals an intriguing equivalence between
two formulations of this theory, one of which is one-dimensional, and the other
of which is infinite-dimensional.  Second, I demonstrate the use of partial
supersymmetry as a tool to obtain the asymptotic energy
levels in non-relativistic quantum mechanics in an exceptionally easy way.
In the end, I discuss possible extensions of this work, including the
possible connections between
partial supersymmetry and renormalization group arguments.
\vskip .5in
\centerline{PACS 03.65.-w, 03.65.Fd, 11.30.Pb}

\Date{10/2003}

\newsec{Introduction}

Supersymmetry is, by now, a familiar mathematical construct for physicists as
well as mathematicians.
It has been invoked not simply as a possible phenomenological symmetry 
of nature, but as a tool with which one can analyze generic, even non-supersymmetric,
physical theories, as well as to derive mathematical results.  Examples of
non-supersymmetric applications of supersymmetry
range from the use of shape invariance to solve exactly soluble quantum 
Hamiltonians \ref\shapeinv{L. E. Gendenshtein, {\it JETP Lett.} 38 (1983) 356\semi 
F. Cooper, A. Khare, and U. Sukhatme, {\it Phys. Rep.} 251
(1995) 267.},  the analysis of monopoles and other topological field
configurations \ref\susysoliton{Z. Hlousek and D. Spector, {\it Nuc.
Phys.} B397 (1993) 173\semi
Z. Hlousek and D. Spector, {\it Nuc. Phys.}
B442 (1995) 413. }, and the calculation of gluon and graviton scattering
amplitudes
\ref\susyamp{S. Parke and T.R. Taylor, {\it Phys. Lett.} 157B (1985) 81;
\hfill \break
F.A. Berends, W.T. Giele, and H. Kuijf, {\it Phys. Lett.} 211B
(1988) 91.
}.

One of the hallmarks of supersymmetry is that it produces structures that afford
exact control over at least some aspects of a physical theory, which in turn
leads to the frequent effectiveness of supersymmetry as an analytical tool when 
a model
can in some fashion be associated with a supersymmetric theory.
Given the extensive usefulness of supersymmetry, any generalizations of 
supersymmetry are also potentially of great interest.  A few years
ago, in work on arithmetic quantum  theories \ref\zetaduality{D. Spector, 
{\it J. Math. Phys.} (1998) 1919.}, I introduced
a notion of partial supersymmetry, in which some, but not all, of the operators
and states of a theory
were paired in supersymmetric fashion, and then explored the use of
partial supersymmetry to develop relations among arithmetic quantum theories.

In the present paper, I return to the subject of partial supersymmetry, and
demonstrate that it has a broader range of applicability.  First, I will
state briefly what is meant by partial supersymmetry.  Then
I will use partial supersymmetry to study a theory with a logarithmic spectrum,
a spectrum which is of interest for its role connecting quantum mechanics
both to number theory \ref\smobius{ 
D. Spector, {\it Commun. Math. Phys. } 127 (1990) 239\semi
B. Julia, ``Statistical Theory of Numbers,'' in {\it Number
Theory and Physics}, edited by J.M. Luck, P. Moussa, and M. Waldschmidt,
{\it Proceedings in Physics } 47 (Springer-Verlag, Berlin, 1990) 276.
    } and to string theory \ref\logtheory{I. Bakas and M. Bowick,  
{\it J. Math. Phys.} 32 (1991) 1881.}.
Partial supersymmetry will reveal the underlying structure of this
theory.  Moving from a specific application to a more general one, I
will use partial supersymmetry
as a tool to extract the asymptotic energy levels of quantum mechanical systems. 
While there are obviously already techniques to obtain asymptotic energy levels,  the
methods described in this section are significant both for their ease of  use -- indeed,
there is probably no easier way to obtain the asymptotic spacing of energy levels in
non-relativistic quantum mechanics -- and for the way 
in which they reveal the symmetry
structure underlying the fact that the asymptotic behavior can be exactly calculated. 
Finally, I discuss some possible interpretations of this symmetry, and consider
additional applications.

\newsec{Partial Supersymmetry}

In this paper, we will for the most part work with one-dimensional
non-relativistic quantum mechanics in the Schr\"odinger
picture.

Consider a Hamiltonian
$H_0$.  The spectrum of this theory will consist of a set of states with
energies $E_0 < E_1 < E_2 < E_3 < \ldots$.  For simplicity, we will consider
the case with a purely discrete spectrum.

Now what happens if we extend this to the supersymmetric case?\foot{While
strictly speaking, one might think of supersymmetry as only the requirement
that $H$ can be written as the square of a hermitian operator, in order for
supersymmetric quantum mechanics to make contact with conventional supersymmetric
field theory, we also require a $Z_2$ grading analogous to $(-1)^F$.}  In this case,
the energies are bounded from below by zero.  The states of positive energy
come in degenerate pairs, while there may also be a supersymmetry singlet
zero energy state.   Thus the
supersymmetric scenario leads to two sectors, either both with the same spectrum, or the
situation in which one sector
has states with energy
$E_0 < E_1 < E_2 < E_3 < \ldots$, and the other sector consists of states
of energy $E_1 < E_2 < E_3 < \ldots$.  Needless to say, the
supersymmetry algebra implies results beyond simply level degeneracies, but
we will not focus on those here.

Now suppose the spectrum has a different structure.  Consider the case
that there are two sectors, one with energies $E_0 < E_1 < E_2 < E_3 < \ldots$,
and the other with energies $E_1 < E_3 < E_5 < \ldots$.  In this case, every
state in the second sector has a partner from the first sector
with degenerate energy, but
only every other state in the original sector has such a partner in
the second sector.  This situation,
in which all of the states in one sector have partners, but only some of the
states in the other sector do, and in which there is a regular pattern to which
states do and do not have partners\foot{Without some regularity, after all, one
should not expect an underlying structure.}, is the situation we shall
term {\it partial supersymmetry}.  In fact, the underlying structure should
have implications beyond degeneracies, e.g., that some operators come in
pairs, and others are singlets, as discussed in \zetaduality, but for the 
purposes of this paper, we will focus on the spectrum.

In the remainder of this paper, we will examine some situations in which partial
supersymmetry can be found and utilized.

\newsec{The Logarithmic Spectrum}

Suppose we have a quantum mechanical Hamiltonian
\eqn\HLdef{H_L = -{\hbar^2\over 2m}{d^2\over dx^2}+V_L(x)~~~}
which has the property that its energy spectrum is purely
logarithmic,
\eqn\logspec{E_n = \epsilon_0 \ln(n), n=1,2,3,\ldots~~~.}
Such a theory has proven of interest in \zetaduality \smobius \logtheory\ 
as a toy model of some features of string theory, in particular
the appearance of a Hagedorn temperature \ref\hagedorn{R. Hagedorn, {\it Nuovo Cim.} 
56A (1968) 1027.}.  This theory and the related ones in which the spectrum is still
logarithmic but there are degeneracies at various levels have also been useful as tools
to draw connections between quantum mechanics and arithmetic number theory, 
invoking, for example, the observation that the physical quantum partition
function for such a theory is the number theoretic Dirichlet series of the 
degeneracy function (with the non-degenerate case yielding the Riemann
zeta function).

Here, our focus is not on the many uses of the logarithmic spectrum,
but rather on the structure of the theory that has
a logarithmic spectrum without degeneracies, and the
way in which partial supersymmetry can reveal that structure.  We will not
in the present section discuss the form of the potential $V_L(x)$, but 
we will come back to that question later in this paper.

Return, then, to consideration of the Hamiltonian $H_L$.  Suppose
we now define a new Hamiltonian, $\tilde H_L = H_L + \epsilon_0 \ln(2)$.
On the one hand, adding a constant has no effect other than to shift
the energies.  On the other hand, in the case at hand, something interesting
arises.

The spectrum of $\tilde H_L$ consists of the energies
\eqn\tildeEn{\tilde E_n = \epsilon_0 \ln(n)+\epsilon_0 \ln(2) 
   = \epsilon_0\ln(2n),\qquad n=1,2,3,\ldots~~~.}
One notes that these levels are degenerate with the alternate
levels of the original Hamiltonian $H_L$.  Consequently,
the combined system, consisting of the levels of $H_L$ and
$\tilde H_L$, is partially supersymmetric.

We can write a Hamiltonian for the combined system very easily.
Introducing fermionic creation and annihilation operators
$f_2^\dagger$ and $f_2$, respectively, which satisfy the standard
anticommutation relations
\eqn\facom{\{f_2,f_2\} = 0 = \{f_2^\dagger,f_2^\dagger\}\qquad
\{f_2^\dagger,f_2\}=1~~~,}
we now define the combination Hamiltonian
\eqn\Hcdefined{H_C = H_L + \epsilon_0\ln(2)f_2^\dagger f_2~~~.}
This theory has two sectors, one with fermion number zero, which has
the logarithmic energy spectrum of $H_L$, and one with fermion number
$+1$, which has the even logarithmic energy spectrum of $\tilde H_L$.

As noted above, $H_C$ has a partial supersymmetry invariance.
Consequently, we expect that within $H_L$, there will be contained
a bosonic term partner to the fermionic addition in \Hcdefined.  Hence
we anticipate that $H_L$ will contain a term of the
form $\epsilon_0\ln(2) b_2^\dagger b_2$, which we shall denote with
the notation
\eqn\HLcontains{H_L \supset \epsilon_0\ln(2) b_2^\dagger b_2~~~,}
where we have introduced the bosonic creation and annihilation operators
satisfying the standard algebra
\eqn\bcom{ [b_2,b_2]=0=[b_2^\dagger,b_2^\dagger]\qquad
[b_2^\dagger,b_2]=1~~~.}

Now there was nothing special about adding energy $\epsilon_0\ln(2)$ to
the original Hamiltonian; one could instead have added $\epsilon_0\ln(3)$, for
example, and then pairing this new system with the original $H_L$, we
would have a system in which every third state of the original Hamiltonian
has a degenerate partner. The associated partial supersymmetry 
in this case would then
indicate that
\eqn\HLcontainiii{H_L \supset \epsilon_0\ln(3) b_3^\dagger b_3~~~.}
Likewise, we would expect that for any other integer
\eqn\HLcontainany{H_L \supset \epsilon_0\ln(j) b_j^\dagger b_j~~~.}

However, this is too much.  For example, once one has included
a term
$\epsilon_0\ln(2) b_2^\dagger b_2$ in $H_L$, this already produces
a state with energy $\epsilon_0\ln(4)$, and so it is overcounting
also to include in $H_L$ a term of the form $\epsilon_0\ln(4) b_4^\dagger b_4$.
Likewise, once $\epsilon_0\ln(2) b_2^\dagger b_2$ and
$\epsilon_0\ln(3) b_3^\dagger b_3$ are both present, this already produces
a state with energy $\epsilon_0\ln(6)$, and so it is overcounting
also to include in $H_L$ a term of the form $\epsilon_0\ln(6) b_6^\dagger b_6$.

This process described above of which bosonic operators to include and
exclude is readily recognized as an implementation of the Sieve of 
Eratosthenes \ref\sieve{Hardy, G.ÊH. and Wright, E.ÊM. {\it An Introduction 
to the Theory of Numbers, 5th ed.} (1979) Oxford, England: Clarendon Press.}, and so the
only terms that should remain in $H_L$ are the ones associated with prime integers. 
Consequently, we conclude that we can re-write the Hamiltonian $H_L$ as
\eqn\HLfactored{
-{\hbar^2\over 2m}{d^2\over dx^2} + V_L(x) 
  = \sum_{k=1}^\infty \epsilon_0 \ln(p_k) b_k^\dagger b_k~~~,}
where $p_k$ is the $k^{th}$ prime, and we have re-labeled the operators
so that $b_k^\dagger$ and $b_k$ are associated with the integer $p_k$
rather than $k$.

While it is clear by
direct calculation that the right side of \HLfactored\ has a logarithmic
spectrum, the above calculation demonstrates {\it why} a theory with
a logarithmic spectrum should have a factorized representation of this
form.  The explanation rests in the series of partial supersymmetries which
can be introduced into this problem.

Perhaps even more strikingly, the equality in \HLfactored\ relates two
quantum mechanical theories: one in $R^1$ and the other in $R^\infty$.
This equivalence of two theories, both bosonic, in such drastically
different dimensions is striking.  A fuller understanding of this phenomenon,
and the factorization to which it is related,
is under investigation.  We note in passing a potential similarity to the
sectorization phenomenon in \ref\digamma{T.J. Allen, C.J. Efthimiou, and D. Spector,
hep-th/0209204}.

\newsec{Asymptotic Energy Levels}

In this section, we will see the broader utility of partial 
supersymmetry by using partial supersymmetry to obtain
the  asymptotic energy
level spacing of quantum mechanical Hamiltonians in a general setting.
While other methods exist to find these levels, 
the partial supersymmetry method introduced here
is exceptionally easy from a calculational point of view, and
obtains its effectiveness from the algebraic structure underlying
the procedure.  
Again, as in the preceding section, we will see that
partial supersymmetry can be used to obtain results about theories
that are not intrinsically supersymmetric.
At the end of this section, we use these asymptotic
methods to return to the question of how the logarithmic spectrum 
of the previous section can
be generated.  

Suppose, then, that we have a theory --- as before, non-relativistic
quantum mechanics in one spatial 
dimension --- with Hamiltonian $H(g)$ and with
an energy spectrum given by $E(g,n)$, $n=1,2,3,\ldots$.  Suppose further
that we
find a transformation $g\rightarrow g^\prime$ such that the energy
levels of the new Hamiltonian $H(g^\prime)$ satisfy $E(g^\prime,n)=E(g,2n)$.
In such a situation, $H(g)$ and $H(g^\prime)$ form a partially supersymmetric
pair.  Now suppose 
further that we find a temperature where the partition functions
of both these Hamiltonians diverge, that is, a Hagedorn temperature.  Then at
such a temperature, we would see a signal of partial supersymmetry, namely
that the ratio of the partition functions is $1/2$, as only the asymptotic
density of states matters at a point where the partition functions diverge.

The key insight is that we can run this argument in reverse.
Suppose we have
a Hamiltonian $H(g)$ for which the partition function diverges at
some temperature $T_H$ (inverse temperature $\beta_H$). 
Then given a transformation of the parameter $g\rightarrow g^\prime$ such
that at temperature $T_H$ the partition functions 
for $H(g)$ and $H(g^\prime)$ both diverge but satisfy
\eqn\Zratio{ {Z(g^\prime,T_H)\over Z(g,T_H)} = {1\over 2}~~~,}
we can conclude that, asymptotically, the density of states of the 
second theory is half that of the first.  This means that the
asymptotic equality
\eqn\Erelation{E(g^\prime,n) \sim E(g,2n)\qquad~~~}
holds, which means that, asymptotically, there is a partial supersymmetry.
Now, in conjunction with some simple dimensional analysis, one can obtain
the asymptotic level spacings in this theory.

What makes this an especially easy method for computing the asymptotic 
spacing of energy levels is that, due to the correspondence
principle, the one necessary calculation --
determination of the (Hagedorn) temperature at which the partition 
function diverges  -- can  be performed in the classical limit, 
as it is the behavior of the system at arbitrarily high energies that
determines if and when the partition function diverges.

We present two examples to demonstrate this method, the second of which will
connect us to the logarithmic spectrum of the previous section.

For the first example, consider the Hamiltonian
\eqn\hamgxr{H = -{\hbar^2\over 2m}{d^2\over dx^2} + gx^r~~~,}
where $g$ and $r$ are positive constants (and we can restrict to the
positive $x$-axis if need be).
We seek to find how the energy levels asymptotically depend on $n$.
To find $\beta_H$, we calculate the classical partition function,
\eqn\Zclassgxr{Z(g,m,\hbar,\beta) ={1\over \hbar} \int dp e^{-\beta p^2/2m}
 \int dx e^{-\beta gx^r}~~~,}
which, it is simple to see, yields
\eqn\Zclassanswer{Z(g,m,\hbar,\beta) 
= {1\over \hbar}\sqrt{2\pi m \over \beta} {c\over (\beta g)^{1/r}}~~~,} 
where $c$ is a dimensionless constant.  Consequently, the partition
function
diverges as $\beta_H \rightarrow 0$ or at infinite temperature.

It is easy to check that for $g^\prime = 2^r g$, the partition functions
satisfy
\eqn\Zratiogxr{
{Z(g^\prime,m,\hbar,\beta_H)\over Z(g,m,\hbar,\beta_H)}={1\over 2}~~~,}
and, consequently, we have that
\eqn\Egxr{ E(2^r g,\hbar,m,n) \sim E(g,\hbar,m,2n)~~~,}
where ``$\sim$'' denotes asymptotic equality.
In other words, the combination of these two theories has partial supersymmetry
in the asymptotic limit. Now it is a simple exercise in dimensional analysis to
see that
\eqn\Egxrdim{E(g,\hbar,m,n) = \bigl({\hbar^2\over m}\bigr)^{r\over r+2} g^{2\over r+2}
f(n)~~~,}
for some function $f(n)$.  Plugging the expression \Egxrdim\ into
the partial supersymmetry relation \Egxr , one readily determines
that $f(n) =n^{2r/(r+2)}$, and thus that the asymptotic spacing
of energy levels is given by
\eqn\Engxr{E(g,\hbar,m,n) 
\sim \bigl({\hbar^2\over m}\bigr)^{r\over r+2} g^{2\over r+2} n^{2r\over r+2}~~~
.}
One easily recognizes the familiar cases $r=2$ (the harmonic
oscillator) and $r\rightarrow\infty$ (the infinite square well).

Note that it is certainly possible to obtain this result through other methods,
such as WKB or Bohr-Sommerfeld quantization, but the method 
presented here is far simpler.  Its power
stems from the fact that we construct a partial supersymmetry that holds in the
asymptotic limit, but since this is the only limit of concern here, this
algebraic property is sufficient
to  obtain the desired result with a minimum of calculation.

We now turn to second calculation of asymptotic level spacing, and in so doing
address the question of what potential $V_L(x)$ in \HLdef\ will lead to a 
logarithmic spectrum.  Because of the nature of our techniques, it will only
address this question asymptotically, but in so doing, will prove that a potential
$V_L(x)$ exists.

Consider, then, a Hamiltonian
\eqn\Hlog{H=-{\hbar^2\over 2m}{d^2\over dx^2} +\epsilon_0 \ln(x) ~~~.}
First, we find the inverse temperature where the partition function for this
Hamiltonian diverges.  One finds
\eqn\Zlog{Z = {1\over\hbar}
 \sqrt{2\pi m\over \beta}\int_{x_0}^\infty dx e^{-\beta\epsilon_0\ln(x)}~~~,}
which is finite for $\beta>1/\epsilon_0$ but diverges at $\beta=1/\epsilon_0$.
(Note the need for an infrared cutoff $x_0$ for this potential.)
Consequently, this theory has a true Hagedorn temperature, a finite temperature
at which the partition function diverges, $T_H=1/\beta_H = \epsilon_0$.

Now consider the transformation $V(x) \rightarrow V(x)+\epsilon_0\ln(2)$.
Under this transformation, 
\eqn\Zmap{Z\rightarrow Z^\prime = {1\over 2^{\beta\epsilon_0}}Z~~~}
at an arbitrary temperature, and thus at the Hagedorn temperature
\eqn\Zlogratio{ {Z^\prime \over Z} = {1\over 2}~~~.}  
Consequently, at
this temperature, we have an asymptotic partial supersymmetry,
and so $E^\prime(n)$ and $E(2n)$ are asymptotically equal.

But the transformation on the potential has an utterly trivial effect
on the energies: $E^\prime(n) = E(n)+\epsilon_0\ln(2)$.  Putting
this all together, we see that, asymptotically,
$E(2n) \sim E(n)+\epsilon_0\ln(2)$.
Note that a similar argument could be made with a shift in the potential by
$\epsilon_0\ln(k)$ for any integer $k>1$.
As a result of this shift identity, one has that
\eqn\Elogasym{E(n) \sim \epsilon_0\ln(n)~~~.}
Therefore, the logarithmic potential $\epsilon_0 \ln(x)$ produces, in
the asymptotic regime, the
logarithmic energy spectrum $\epsilon_0 \ln(n)$, the spectrum
that we studied in
the previous section.

We make two observations here.  First, this result for the logarithmic
potential can be re-obtained
and verified by standard methods.  Second, we
note that the asymptotic argument in this case is enough to 
ensure that there is some potential that will yield exactly 
the logarithmic spectrum.  By modifying the small $x$ behavior
of the potential in \Hlog, one can get the lower energies to
match up with the spectrum \logspec, while as long as the asymptotic
behavior of the potential at large $x$ is unchanged, the spectrum
will remain logarithmic at large $n$.  Thus we have come full circle,
using the partial supersymmetry analysis of asymptotic energy levels 
in quantum mechanics to shed light in turn on the theory with a logarithmic
spectrum, the theory with which we started exploring the consequences of
partial supersymmetry.

\newsec{Conclusions}

In this paper, I have demonstrated the usefulness of partial supersymmetry as
a tool for the analysis of quantum mechanical systems.  We have
seen the kind of energy spectra that are the hallmark of partial supersymmetry.
We have also seen that we can 
enhance theories so that they have such a spectrum, use the 
partial supersymmetry of the enhanced theory to analyze the original theory,
and thus obtain results about theories that had no supersymmetry, partial
or otherwise, to begin with.

Our two applications in this paper have been to understand the factorization
into creation and annihilation operators that occurs when there is a
purely logarithmic spectrum, and to obtain the asymptotic level spacings
in non-relativistic quantum mechanics with only the simplest of calculations,
since partial supersymmetry enables us to focus specifically on the asymptotic
behavior of the spectrum without distractions.
Note that partial supersymmetry shed light
on the logarithmic case as well when using these asymptotic techniques.

There are two extensions under current investigation.
One is the extension of these techniques 
to operators other than the Schr\"odinger
operator of one-dimensional quantum mechanics.  The other is the
exploitation of algebraic aspects of partial supersymmetry at the operator
level, not simply at the level of energy spectra.

In addition, it is important to note that there is an
interpretation of this work that bears consideration.  Notice that the
section on the logarithmic spectrum and the section on the logarithmic
potential (which asymptotically has a logarithmic spectrum)
both involve a common
transformation,
$H\rightarrow H+\epsilon_0\ln(2)$.

In the first case, the shift in the energy leads us to thin out the energy
levels, removing every other one (though of course we know, also, that this
transformation gives the original theory at shifted energies).  Thus it 
represents a kind of renormalization group transformation, where we map every
two states into one.  In the second case, the shift in the energy can be understood
as a shift in the infrared cutoff (consider the partition function \Zlog ).
Again, a shift in the infrared cutoff should ultimately have no physical effect
on the theory, and understanding how quantities are affected by this shift in
the cutoff is the job of the renormalization group.  Thus we see that there is
an intimate connection between the applications of partial supersymmetry 
considered here and the
renormalization group, and understanding better this connection and its interpretation
is work currently underway.

Finally, should we have been surprised that a supersymmetric method can be
used to obtain asymptotic level spacings?  I would argue not.  It is well-known
that supersymmetry can be used to obtain the energy levels in the known
exactly soluble systems.  The asymptotic limit of the quantum systems in 
question
is exactly soluble (that is, the asymptotic behavior can be determined
exactly to leading order), and thus it is reasonable to expect that there 
would be a supersymmetric way to find this asymptotic behavior.
It turns out, however, that it is partial supersymmetry that is actually
the relevant tool.

I thank Ted Allen for conversations.

\listrefs
\bye